# Spin susceptibilities in zigzag graphene nanoribbons

Juan Antonio  Casao Pérez

*Dpto. Ingeniería Electrónica y Comunicaciones. Escuela de Ingeniería y Arquitectura.*
*Universidad de Zaragoza. María de Luna 3, 50012 Zaragoza. Spain.  [casao@unizar.es](mailto:casao@unizar.es)*

**Abstract.**

The simple Hubbard Hamiltonian with the mean field approximation is used to know about the energy bands and spin susceptibilities of a zigzag graphene nanoribbons. Depending on the electron doping antiferromagnetic or ferromagnetic configurations are possible; in the former, an energy gap exits which is proportional to the Hubbard parameter, while in the latter the up and down spin bands intersect the Fermi level.  Due to the two dimensional nature of the system a square susceptibility matrix is necessary to explain the self-correlations and correlations between spins in the ribbon unit cell. The transverse spin susceptibilities are computed and the static case for ferromagnetic solutions is examined, as a function of the electron doping.

## 1. Introduction.

The analysis of the electronic properties of graphene and the systems based on it, has propelled a great amount of research on such subjects [1-2]. Particularly, the energy bands and magnetic properties of zigzag graphene nanoribbons (ZGNR) have been widely explored using the DFT [3-4], the hybrid exchange density functional method [5] and the tight-binding Hubbard Hamiltonian [6-8]. In [7], within the mean field approximation, the energy per ribbon unit cell is used to show the phase diagram of such a system. A transition from an antiferromagnetic (AF) configuration to a ferromagnetic (F) state when the electron doping is increased, has been established. In [9] the orbital diamagnetic and the Pauli paramagnetic susceptibilities of ZGNRs are computed using a single orbital tight-binding model, without considering the Coulomb interaction. In this article, the simple Hubbard Hamiltonian in the mean field approximation is used to analyze the energy levels and the spin susceptibilities of ZGNRs. Due to the two-dimensional nature of the ZGNR, these correlations emerge forming a square matrix whose dimension is the number of atoms in the unit cell of the ribbon. On and off diagonal matrix elements are examined for F solutions, as a function of doping.

## 2. Theoretical model.

Analytical studies of GNRs use the Dirac equation without the Coulomb interaction [10]. Here, we consider a ZGNR with translational symmetry along the x axis, formed with $N_z$ (even) longitudinal chains; as shown in Fig. 1. The simple Hubbard Hamiltonian using the mean field approximation can be written as

$$
\begin{aligned}
H = & -t \cdot \sum_i \left\{ \sum_{j=1,3,5,\dots;\sigma}^{N_z-1} \left[ \left( a_{ij\sigma}^+ b_{ij\sigma} + a_{ij\sigma}^+ b_{i,j-1,\sigma} + a_{ij\sigma}^+ b_{i-1,j-1,\sigma} \right) + h.c. \right] - \right. \\
& t \cdot \sum_{j=2,4,6,\dots;\sigma}^{N_z-2} \left[ \left( a_{ij\sigma}^+ b_{ij\sigma} + a_{ij\sigma}^+ b_{i+1,j-1,\sigma} + a_{ij\sigma}^+ b_{i,j-1,\sigma} \right) + h.c. \right] - \\
& \left. t \cdot \sum_\sigma \left[ \left( a_{i,N_z,\sigma}^+ b_{i+1,N_z-1,\sigma} + a_{i,N_z,\sigma}^+ b_{i,N_z-1,\sigma} \right) + h.c. \right] + U \cdot \sum_{j,l=A,B} \left[ n_{ijl\uparrow} \cdot \left\langle n_{ijl\downarrow} \right\rangle + n_{ijl\downarrow} \cdot \left\langle n_{ijl\uparrow} \right\rangle \right] \right\}
\end{aligned}
$$

(1)

This model has been extensively used to study ground-state properties and magnetic correlations in graphene systems [6, 7, 8, 11, 12, 13]. A Fourier transformation to the momentum space $k_x$ is done, using the relations $c_{ij\sigma}^+ = 1/\sqrt{N} \cdot \sum_{kx} e^{-ikx \cdot Xi} c_{kx,j\sigma}^+$, $c=a$ or $b$ for A or B atoms respectively. Then, $\left\langle n_{jA\downarrow} \right\rangle = (1/N) \cdot \sum_{kx} \left\langle n_{kx,jA\downarrow} \right\rangle = (1/N) \cdot \sum_{kx} \left\langle a_{kx,j\downarrow}^+ a_{kx,j\downarrow} \right\rangle$ represents the average number of electrons with spin down at the j-th A atom of the conventional cell of the nanoribbon; which contains $2N_z$ atoms, as shown with dashed lines in Fig. 1 (and similar expressions for the rest of the number operators). Taking into account that $c_{kx,j\sigma}^{+}|0\rangle = |c,k_x,j,\sigma\rangle$, $c=a, b$; with $k_x$ and $\sigma$ fixed, the independent particle Hamiltonian $H_{1p}(k_x,\sigma)$ of Ref. [14] is obtained; with the U-elements in the diagonal included. An initial conjecture for $\langle n_{0B\downarrow}\rangle, \langle n_{1A\downarrow}\rangle, \langle n_{1B\downarrow}\rangle, ..., \langle n_{NzA\downarrow}\rangle$ is given. Then, the eigenenergies $E(k_x,k_y,\sigma)$ and eigenvectors $|k_x,k_y,\sigma\rangle$ of $H_{1p}(k_x,\sigma)$ are computed; that is

$$H_{1p}(k_x,\sigma)|k_x,k_y,\sigma\rangle = E(k_x,k_y,\sigma)|k_x,k_y,\sigma\rangle \tag{2a}$$

$$|k_x,k_y,\sigma\rangle = \sum_{j=1}^{Nz} \alpha(k_x,k_y,j,\sigma)\cdot|a,k_x,j,\sigma\rangle + \beta(k_x,k_y,j-1,\sigma)\cdot|b,k_x,j-1,\sigma\rangle \tag{2b}$$

where $k_y$ can take $2N_z$ possible values. At T=0 K, it can be shown that (j=1,2,…,$N_z$)

$$\langle n_{jA\downarrow}\rangle = \int_0^1 d\bar{k}_x \sum_{ky}^{occupied} \left|\alpha(\bar{k}_x,k_y,j,\downarrow)\right|^2 \tag{3}$$

with $\bar{k}_x = k_x(a/\pi)$ the normalized wave vector; and analog equations for $\langle n_{jA\uparrow}\rangle$, $\langle n_{jB\downarrow}\rangle$ and $\langle n_{jB\uparrow}\rangle$ [8, 15]. These equations are solved self-consistently until convergence is reached. In the procedure, the charge neutrality condition on every atom in the cell of the ribbon is assumed; that is, $(n_{jl\uparrow})+(n_{jl\downarrow})=n_{e-ato}$, $l$=B, $j$=0,1,…, $N_z$-1 and $l$=A, $j$=1,2,…,$N_z$ ; where $n_{e-ato}$ is the number of electrons per atom [12]. At half filling ($n_{e-ato}$=1.0), depending on whether the initial solution is symmetric or not, a ferromagnetic (F) or antiferromagnetic (AF) solution is reached respectively. Once a solution has been achieved, the simple particle Hamiltonian can be expressed by

$$H_{1p} = \sum_{kx,\sigma} H_{1p}(k_x,\sigma) = \sum_{kx,\sigma}\sum_{ky} E(k_x,k_y,\sigma)\cdot|k_x,k_y,\sigma\rangle\langle k_x,k_y,\sigma| \tag{4}$$

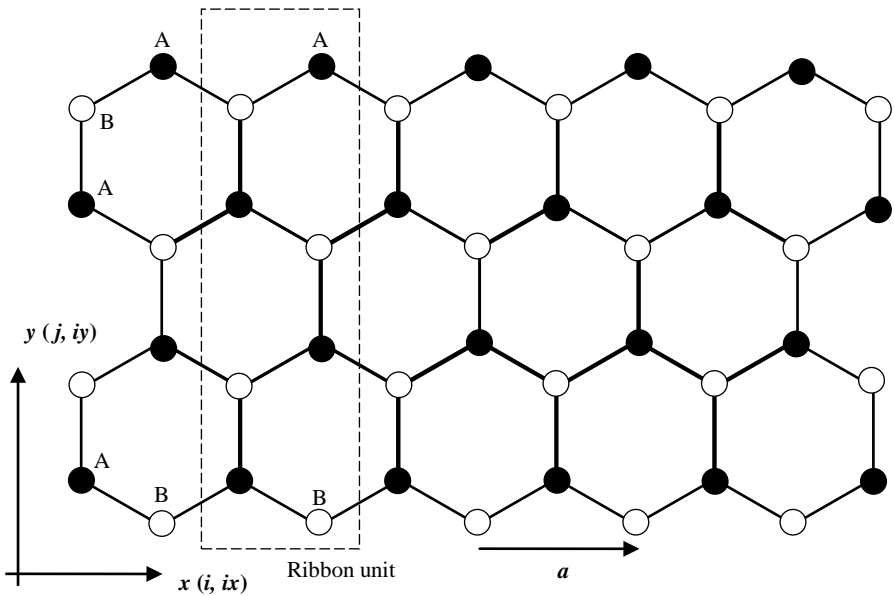

**Fig. 1. ZGNR with $N_z=4$ chains and width $W=(\sqrt{3}/2)aN_z$.**

At $k_x=\pi$, the Hamiltonian $H_{1p}(k_x,\sigma)$ can be analytically solved, and the edge state energies are $E_{kx=\pi/2,L,\uparrow}=U\cdot\langle n_{0B\downarrow}\rangle$, $E_{\pi/2,R,\uparrow}=U\cdot\langle n_{NzA\downarrow}\rangle$ for the left (L) and right (R) up levels and, $E_{\pi/2,L,\downarrow}=U\cdot\langle n_{0B\uparrow}\rangle$, $E_{\pi/2,R,\downarrow}=U\cdot\langle n_{NzA\uparrow}\rangle$ for the left (L) and right (R) down levels. From that, the three important configurations are: the no-magnetic solution, with $E_{kx,L,\uparrow}=E_{kx,L,\downarrow}$, $E_{kx,R,\uparrow}=E_{kx,R,\downarrow}$ ; secondly, the AF solution, $E_{kx,L,\uparrow}=E_{kx,R,\downarrow}$ and $E_{kx,L,\downarrow}=E_{kx,R,\uparrow}$ (Fig. 2a); and thirdly, the F solution, $E_{kx,L,\uparrow}=E_{kx,R,\uparrow}$ and $E_{kx,L,\downarrow}=E_{kx,R,\downarrow}$ (Fig. 2b). Simulations with $N_z$=8, $t$=1.0 eV (normalized), U/$t$=1.2 and, with the $k_x$ space (0• $k_x$• π/a) discretized in 1601 points; are shown in Fig. 2. In Fig. 2a, a small electron doping such as $n_{e-ato}$=1.001875 is assumed and, the energy bands E↑=E↓ are shown. A magnetic moment at the left edge $m_L = n_{0B\uparrow} - n_{0B\downarrow} = 0.286 = -m_R = -\left(n_{NzA\uparrow} - n_{NzA\downarrow}\right)$ (in units of $\mu_B$) is obtained. The electron-electron interaction is the responsible of producing an energy gap, which is given by $E_g = E_{R\uparrow} - E_{L\uparrow} = U \cdot \left[n_{NzA\downarrow} - n_{0B\downarrow}\right]$ [16]. In Fig. 2b, $n_{e\text{-}ato}$=1.00625 and, from different initial conditions, the F solution is obtained. Now $m_L = m_R$=0.277, $E_{kx,L,\uparrow} = U \cdot 0.365$ , $E_{kx,L,\downarrow} = U \cdot 0.642$ , and the Fermi level (~0.63) intercepts the upper edge spin-up band and the lower edge spin-down band (see Fig. 2b); an essential point to explain instabilities in the ferromagnetic state, as commented below.

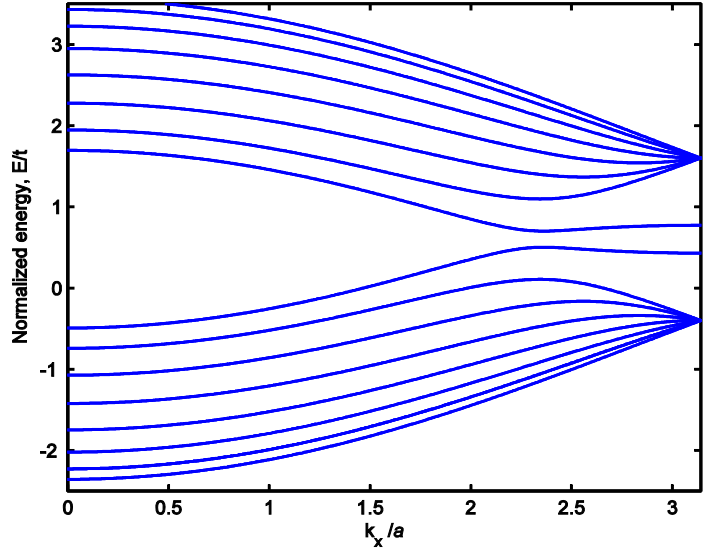

*(2a)*

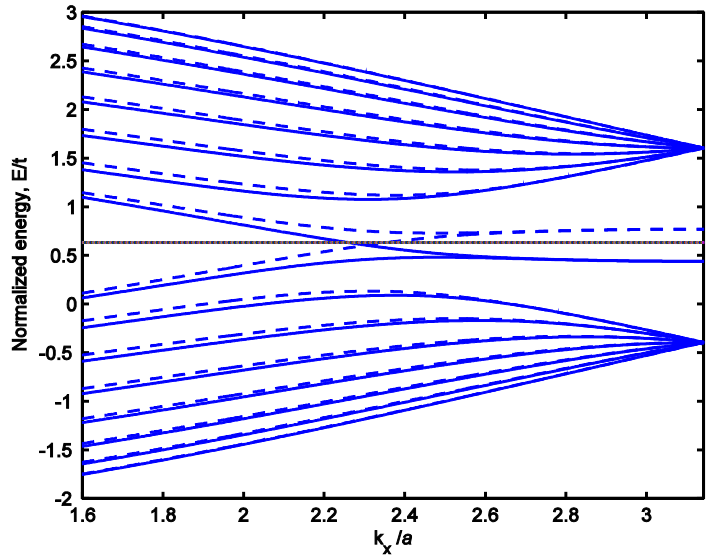

*(2b)*

**Figure 2.** **Energy bands of a ZGNR with $N_z$=8. By symmetry $E(k_x)$=$E(-k_x)$. a) Antiferromagnetic solution with an electron doping per ribbon cell $\delta n$=0.03 . b) Ferromagnetic solution. $\delta n$=0.1 . $\uparrow$ bands, continuous line; $\downarrow$ bands, dashed lines.**

### 3. Spin susceptibilities.

We consider a material system with a spin density $\vec{S}(\vec{r}) = \Psi^+(\vec{r}) \cdot \vec{s} \cdot \Psi(\vec{r})$ , which is perturbed by an external magnetic field $B_{ext}^{\alpha}(\vec{r},t)$, $\alpha = x,y,z$; so that the perturbing Hamiltonian is given by

$$H_t^{/} = -\frac{g\mu_B}{\hbar} \cdot \sum_{\alpha} \int d\vec{r} \, S^{\alpha}(\vec{r}) \cdot B_{ext}^{\alpha}(\vec{r},t) \tag{5}$$

By means of the theory of linear response [17, 18], the expected value of the perturbation in the operator $A(\vec{r},t)$ can be written as

$$\Delta \langle A(\vec{r},t) \rangle = \langle A \rangle_{H+H'} - \langle A \rangle_H = \frac{-i}{\hbar} \cdot \int_{-\infty}^{t} dt' \left\langle \left[ A^H(\vec{r},t), H_{t'}^{/H}(t') \right]_- \right\rangle_H \tag{6}$$

where $A^H(\vec{r},t)$ is the Heisenberg representation, with H the Hamiltonian without the perturbation. In $H_{t'}^{/H}(t')$ the time dependence $t'$ comes from one side, because of an explicit time dependence of the external field and, from the other side, because of the time evolution imposed by the Heisenberg law, of the coupled operator to the external field; $\vec{S}(\vec{r})$ in this case. Our system is the ribbon, whose unit cell can be modeled as a discrete set of $2N_z$ spins, one per each atom; that is, $(\vec{r} \rightarrow i_x, i_y)$ ; so that $S^x(i_x, i_y, A) = (\hbar/2) \cdot \left( a_{ix,iy\downarrow}^+ a_{ix,iy\uparrow} + a_{ix,iy\uparrow}^+ a_{ix,iy\downarrow} \right)$, is the x component of the localized spin in the A atom at position $i_x$, $i_y$, with $i_y=1,2,\ \ ,N_z$ . Therefore, whichever atom is (A or B) with $i_y$, $j_y$ =1, 2,…, Nat=$2N_z$ , applying Eq. (6) to the operator $S^{\beta}(i_x, i_y, t)$ , $\beta = x,y,z$ , we have

$$\Delta \langle S^{\beta}(i_x, i_y; t) \rangle = \frac{aW}{Nat} \cdot \frac{i}{\hbar} \cdot \frac{g\mu_B}{\hbar} \cdot \sum_{\alpha} \sum_{jx,jy} \int_{-\infty}^{t} dt' \, \left\langle \left[ S^{\beta}(i_x, i_y; t), S^{\alpha}(j_x, j_y; t') \right]_- \right\rangle \cdot B_{ext}^{\alpha}(j_x, j_y; t') \tag{7}$$

Because of the translational symmetry along x axis, it is convenient to work in the transformed domain $q_x$ ; so ($\alpha$=x,y,z)

$$S^{\alpha}(q_x, i_y A) = S_{iyA}^{\alpha}(q_x) = \frac{1}{\sqrt{N}} \sum_{ix} e^{-iqx \cdot Xix} \cdot S^{\alpha}(i_x, i_y A) \tag{8}$$

and $i_y A$=1,2,…,$N_z$, for A atoms. With $B_{ext}^{\alpha}(i_x, i_y; t) = B_{ext}^{\alpha}(i_y, t) \cdot e^{iqx \cdot Xix}$, we arrived at

$$\Delta \langle S^{\beta}(q_x, i_y, ; t) \rangle \equiv \frac{1}{N} \sum_{ix} e^{-i \cdot qx \cdot Xix} \cdot \Delta \langle S^{\beta}(i_x, i_y; t) \rangle \ =$$

$$= \frac{aW}{Nat} \cdot \frac{i}{\hbar} \cdot \frac{g\mu_B}{\hbar} \sum_{\alpha, jy}^{x,y,z; Nat} \int_{-\infty}^{t} dt' \, \left\langle \left[ S^{\beta}(q_x, i_y; t), S^{\alpha}(-q_x, j_y; t') \right]_- \right\rangle \cdot B_{ext}^{\alpha}(j_y; t') \tag{9}$$

It is assumed that our system is translation-invariant time; consequently the time dependence with $t$ and $t'$ in the commutator will be in the form $t$-$t'$ . Therefore, we take the Fourier transform to the frequency domain $\omega$ of Eq. (9), and obtain

$$\Delta\left\langle S^{\beta}(q_x, i_y; \omega)\right\rangle = \int_{-\infty}^{\infty} dt\, e^{i\omega t} \cdot \Delta\left\langle S^{\beta}(q_x, i_y; t)\right\rangle = aW \cdot \frac{g\mu_B}{\hbar^2} \cdot \sum_{\alpha}^{x,y,z} \sum_{jy}^{Nat} \chi_{iy,jy}^{\beta\alpha}(q_x; \omega) \cdot B_{ext}^{\alpha}(j_y; \omega)$$

(10)

$$\chi_{iy,jy}^{\beta\alpha}(q_x; \omega) = -\frac{1}{Nat} \cdot (-i) \int_{-\infty}^{\infty} dt\, e^{i\omega t} \cdot \theta(t) \cdot \left\langle \left[ S^{\beta}(q_x, i_y; t), S^{\alpha}(-q_x, j_y; 0) \right]_{-} \right\rangle$$

(11)

Since $S_{ix,iy}^x$ , $S_{ix,iy}^y$ are linear combinations of $S_{ix,iy}^+$ and $S_{ix,iy}^-$ and, because of the spin rotational symmetry of H; for magnetic fields applied parallel to the lattice plane, we are interested in the transverse spin susceptibility ( $i_y$ , $j_y$ =1, 2,…, $N_z$ , A atoms), [19, 20, 21]

$$\chi_{iyA,jyA}^{+-}(q_x; t) = -\frac{1}{Nat} (-i)\theta(t) \cdot \left\langle \left[ S_{iyA}^+(q_x; t), S_{jyA}^-(-q_x; 0) \right]_{-} \right\rangle$$

(12)

where $S_{iyA}^+(q_x; t) = \hbar/\sqrt{N} \cdot \sum_{ix} e^{-iqx \cdot Xix} \cdot a_{ix,iy\uparrow}^+(t)\, a_{ix,iy\downarrow}(t) = \hbar/\sqrt{N} \cdot \sum_{kx} a_{kx,iy\uparrow}^+(t)\, a_{kx+qx,iy\downarrow}(t)$ .

In order to compute $\chi_{iyA,jyA}^{+-}(q_x; t)$ , we consider that the n-particle Hamiltonian H can be written as $H = \sum_{kx,ky,\sigma} E(k_x, k_y, \sigma) \cdot \gamma_{kx,ky,\sigma}^+ \gamma_{kx,ky,\sigma}$ , with

$$\gamma_{kx,ky,\sigma}^+(t) = \sum_{j=1}^{Nz} \alpha(k_x, k_y, j, \sigma) \cdot a_{kx,j\sigma}^+ + \beta(k_x, k_y, j-1, \sigma) \cdot b_{kx,j-1\sigma}^+$$

(13)

Then, the time evolution of the $\gamma$ operators is $\gamma_{kx,ky,\sigma}^+(t) = e^{iHt/\hbar} \cdot \gamma_{kx,ky,\sigma}^+(t=0) \cdot e^{-iHt/\hbar} = e^{iE(kx,ky,\sigma)t/\hbar} \cdot \gamma_{kx,ky,\sigma}^+(0)$ . Using the inverse transformation of Eq.(13) and the relation $\left[ c_\nu^+ c_\mu, c_{\nu'}^+ c_{\mu'} \right] = c_\nu^+ c_{\mu'} \cdot \delta_{\mu,\nu'} - c_{\nu'}^+ c_\mu \cdot \delta_{\mu',\nu}$ [17], we attain

$$\left\langle \left[ S_{iyA}^+(q_x, :t), S_{jyA}^-(-q_x; 0) \right]_{-} \right\rangle = \frac{1}{N} \cdot \sum_{kx} \sum_{ky,ky'} e^{i\left( E(kx,ky,\uparrow) - E(kx+qx,ky',\downarrow)\right)t/\hbar} \times$$

$$\alpha^*(k_x, k_y, i_y, \uparrow) \cdot \alpha(k_x+q_x, k_y', i_y, \downarrow) \cdot \alpha^*(k_x+q_x, k_y', j_y, \downarrow) \cdot \alpha(k_x, k_y, j_y, \uparrow) \times \qquad (14)$$

$$\left[ n_{FD}\left(E(k_x, k_y, \uparrow)\right) - n_{FD}\left(E(k_x+q_x, k_y', \downarrow)\right) \right]$$

where $n_{FD}\left[E(k_x, k_y, \sigma), E_F, T\right]$ is the Fermi Dirac function.  Its Fourier transform will be given by

$$\chi_{iyA,jyA}^{+-}(q_x, \omega+i\eta) = -\frac{1}{Nat}(-i)\int_{0}^{\infty} dt\, e^{i(\omega+i\eta)t} \cdot \left\langle \left[ S_{iyA}^+(q_x; t), S_{jyA}^-(-q_x; 0) \right]_{-} \right\rangle =$$

$$= -\frac{1}{N \cdot Nat} \cdot \sum_{k_x} \sum_{k_y, k_y'} \alpha^*(k_x, k_y, i_y, \uparrow) \cdot \alpha(k_x + q_x, k_y', i_y, \downarrow) \alpha^*(k_x + q_x, k_y', j_y, \downarrow) \cdot$$

$$\cdot \alpha(k_x, k_y, j_y, \uparrow) \times \frac{\left[ n_{FD}(E(k_x, k_y, \uparrow)) - n_{FD}(E(k_x + q_x, k_y', \downarrow)) \right]}{\omega + i\eta + E(k_x, k_y, \uparrow) - E(k_x + q_x, k_y', \downarrow)}$$

(15)

Each element in the summatory of Eq. (15) can be seen as the product of two factors; the first one, which is formed by the four wave amplitudes, takes into account the structure of the eigenstates; whereas the second, looks into the energy levels and their occupations. Similarly, the correlations between a spin in A atom at $i_y$ and another spin in B atom at $j_y$, between a spin in B atom at $i_y$ and another one in A atom at $j_y$ and, finally, between a spin in B atom at $i_y$ and another one in B atom at $j_y$; can be computed. Therefore, we arrive at the following (Nat×Nat) spin susceptibility matrix

$$\chi^{+-}(q_x, \omega + i\eta) = \begin{pmatrix} \chi_{0B,0B}^{+-} & \chi_{0B,1A}^{+-} & \chi_{0B,1B}^{+-} & \cdots & \chi_{0B,NzA}^{+-} \\ \chi_{1A,0B}^{+-} & \chi_{1A,1A}^{+-} & \chi_{1A,1B}^{+-} & \cdots & \chi_{1A,NzA}^{+-} \\ \chi_{1B,0B}^{+-} & \chi_{1B,1A}^{+-} & \chi_{1B,1B}^{+-} & \cdots & \chi_{1B,NzA}^{+-} \\ \cdots & \cdots & \cdots & \cdots & \cdots \\ \chi_{NzA,0B}^{+-} & \chi_{NzA,1A}^{+-} & \chi_{NzA,1B}^{+-} & \cdots & \chi_{NzA,NzA}^{+-} \end{pmatrix}$$

(16)

When $N_z$ is even and in the ferromagnetic solution, there is symmetry around the ribbon center and obviously, $\chi_{0B,0B}^{+-} = \chi_{NzA,NzA}^{+-}$, $\chi_{1A,1A}^{+-} = \chi_{Nz-1B,Nz-1B}^{+-}$ ....; $\chi_{0B,1A}^{+-} = \chi_{NzA,Nz-1B}^{+-}$, ...; so that the unknowns reduce notably.

## 4. Results and discussion.

In order to study the stability of the spin system, first we analyze the static transverse susceptibility as a function of the wave vector $q_x$. In Fig. 3a that self-correlation function at the edge of the ribbon $\chi_{0B,0B}^{+-}(q_x, \omega = 0)$ is shown (arbitrary scale) for three different concentrations of electron per atom, $n_{e\_ato}$=1.00625, 1.025 and 1.05; which correspond to ferromagnetic solutions. In the three curves there are peaks at values of $q_x$ greater than 0, showing that a spin density wave with that finite value of $q_x$ is a possible stable solution [21]. The origin of such a peak can be understood if we focus on the energy bands of the edge states in Fig. 3b (for $n_{e\_ato}$=1.0025), and take into account the last factor in Eq. (15). The Fermi level, $E_{FD}$~0.71, crosses the spin-up edge band at $k_{xFD1}$ and the spin-down edge band at $k_{xFD2}$. When $k_x$ is approaching $k_{xFD1}$ from the left and $\delta q_x = k_{XFD2} - k_{xFD1}$; $E(k_x, k_y, \uparrow)$ is not occupied, $E(k_x + \delta q_x, k_y', \downarrow)$ is occupied and the difference between them comes close to 0. Just immediately after coming by $k_{xFD1}$, the occupation indexes change but the denominator is still next to zero; and these two contributions add up to get a maximum. Because of an increase in $n_{e\_ato}$ produces an increase in the Fermi level, the difference ($k_{xFD2}$-$k_{xFD1}$) grows up and the maximum moves to right. Also, a second peak is appreciated clearly when $n_{e\_ato}$=1.0025 . Since $E(-k_x)=E(k_x)$ and with the band structure displaced into the interval $[\pi, 2\pi]$, the Fermi level crosses the lower spin-down edge band at $k_{xFD3}$, with $k_{xFD3} = 2\pi - k_{xFD2}$ . Then, when $k_x$ is approaching $k_{xFD1}$ from the left and $\delta q_x$ is slightly greater than $k_{xFD3}$-$k_{xFD1}$ , the energy difference between the up and down levels is close to zero; consequently a maximum is produced. In Fig. 3c, the Fermi level (~0.73) intersects the two spin-down edge bands; originating the two maxima obtained, instead of just only one. It is important to note that these instabilities can be explained using only the edge bands. These bands exist for $k_x$ greater than $2\pi/3$ [9, 10, 11], and for such a $k_x$ the wave amplitudes at y=0 ($i_y$=0, $N_z$) $\beta(k_x, k_y, 0, \uparrow)$ and $\beta(k_x + \delta q_x, k_y', 0, \downarrow)$ take significant values.

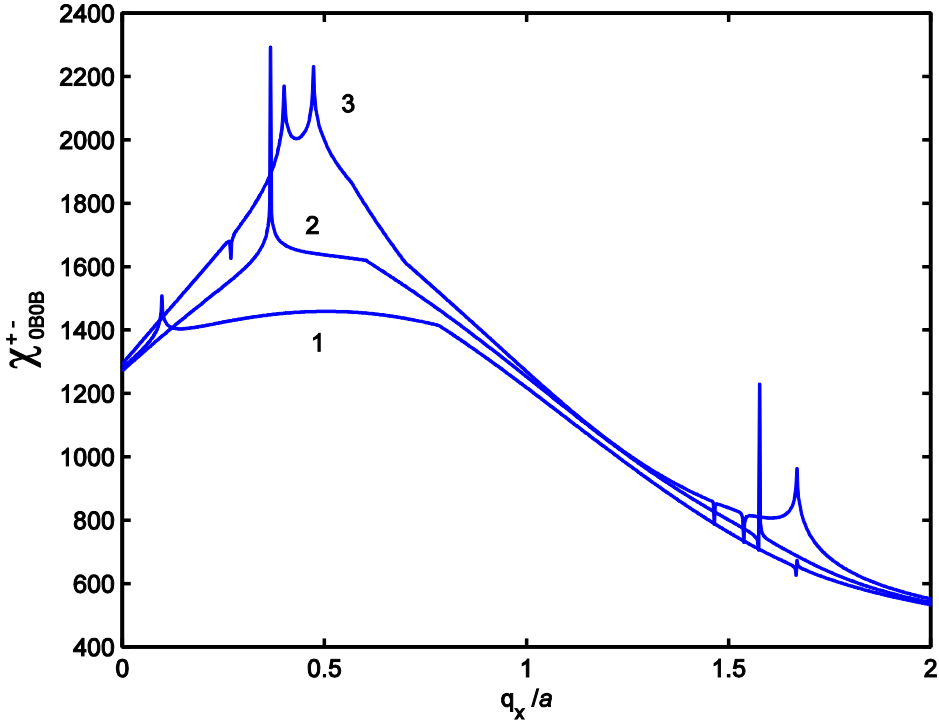

**(3a)**

**Fig. 3.** **(a) Diagonal spin susceptibility at atom 0B, for $n_{e\_ato}$=1.00625(1), 1.025(2) and 1.05(3). (b) Spin-up edge bands (continuous) and spin-down edge levels (dashed) for (2). (c) Spin-up edge bands (continuous) and spin-down edge levels (dashed) for (3).**

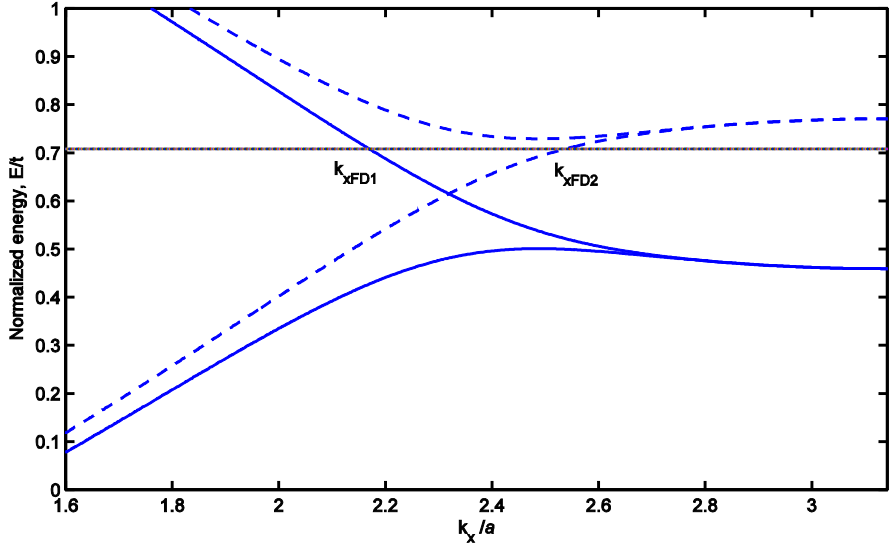

*(3b)*

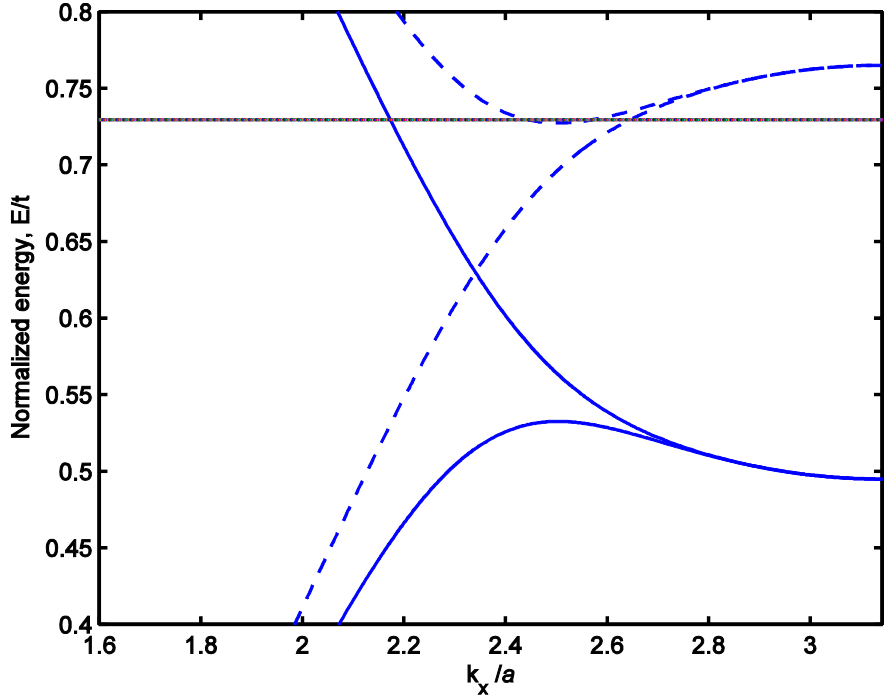

*(3c)*

Also, other elements on and off the diagonal of Eq. (16) have been analyzed. Since the origin of the peaks is the energy factor, all matrix elements show them. However, because of the modulating effect of the wave amplitudes at different sites of the ribbon unit cell, in some susceptibilities such peaks appear to be missing. This is the case of $\chi^{+-}_{1A,1A}(q_x, \omega = 0)$. In Fig. 4 (arbitrary scale), $\chi^{+-}_{1B,1B}(q_x, \omega = 0)$, $\chi^{+-}_{0B,1A}(q_x, \omega = 0)$ and $\chi^{+-}_{0B,1B}(q_x, \omega = 0)$ are shown for $n_{e\_ato} = 1.00625$ in (a) and, $n_{e\_ato} = 1.0025$ in (b). In Eq. (10), it is important to note that a sum extended to all sites of the ribbon unit cell has to be performed. Therefore, the perturbation in the magnetic

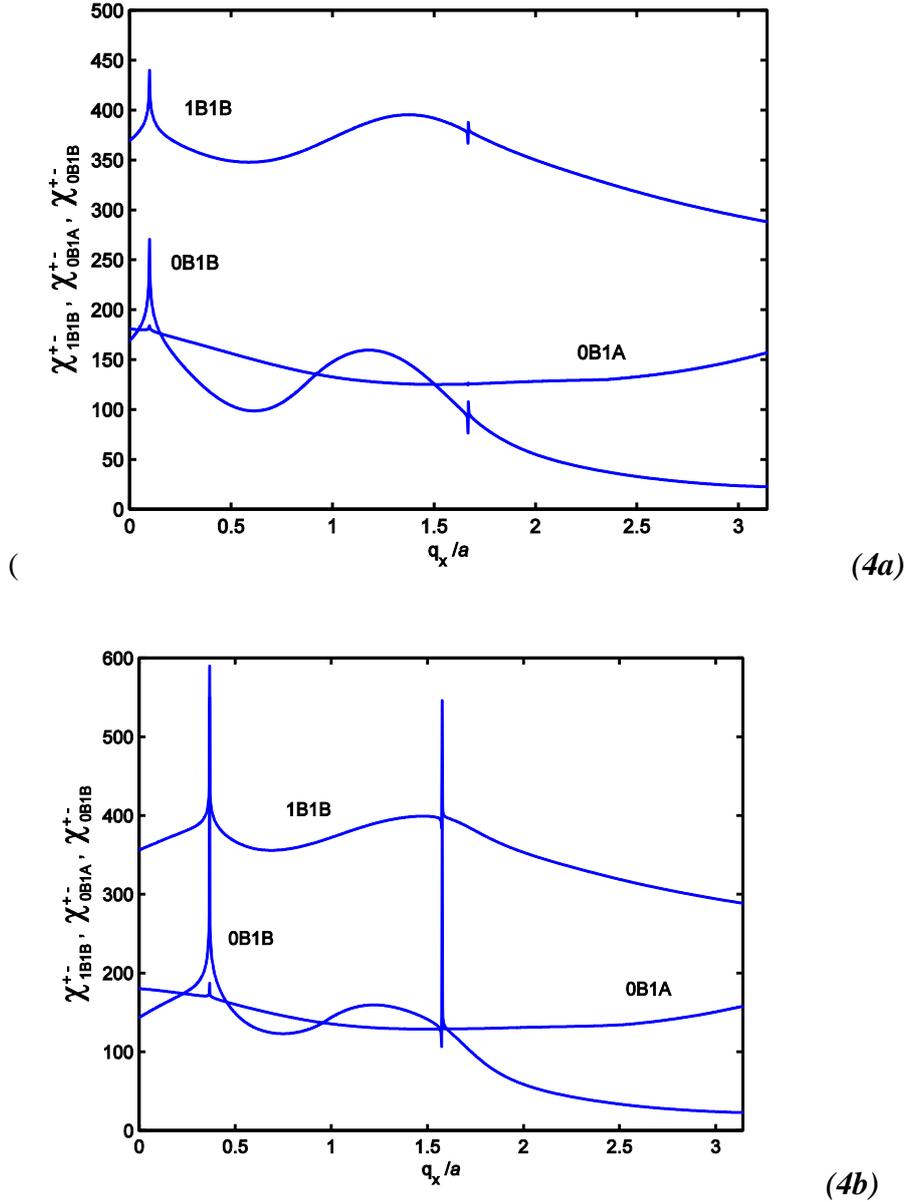

( **(4a)**

**(4b)**

*Figure 4. (a) Spin self-correlation in atom 1B and spin correlations between atoms 0B and 1A; and between 0B and 1B. (b) The same as before but $n_{e\_ato}=1.0025$ .*

moment at atom 0B can be enhanced due to a magnetic field in other atoms; specially for $q_x$ taking the peak values. These figures show that this effect depends on the doping level.

In these results, the temperature has been assumed equal to 0 and defect-free edges are considered. Nevertheless, the effect of the temperature and different kinds of edge terminations can be easily incorporated, in order to study their influence on the stability of such edge magnetism [4]. Also, the Hamiltonian must be modified significantly if we are to analyze possible non-collinear solutions [22, 23]. On the other hand, up to now only the static ($\omega=0$) susceptibilities have been considered; although Eq. (15) give us the dynamical susceptibilities. However, it is necessary to handle the Coulomb interaction using the random phase approximation, to know about the low energy spin wave excitations. [20, 21, 24].

## 5. Conclusions.

In summary, the simple Hubbard model has been used to study the energy bands and spin susceptibilities of a ZGNR. Depending on the electron doping an AF or F solution is obtained. Since the ribbon unit cell contains $2N_z$ spins, self-correlations and correlations between different spins must be taken into account. Static susceptibilities show that there are certain values of $q_x$ at which an instability can appear. Finally, these phenomena have been proved to be strongly dependent on the electron doping.

**Acknowledgements**.

This work is supported by The Science and Innovation Minister through the project FIS2009-12648-C03-02.